\documentclass[twocolumn,aps,prd,epsf]{revtex4}
\usepackage{epsfig}
\usepackage{color}

\begin{document}
\title{ 
First study of the gluon-quark-antiquark static potential in SU(3) Lattice QCD
}
\author{
P. Bicudo}
%\email{bicudo@ist.utl.pt}
\author{
M. Cardoso}
%\email{mjdcc@cftp.ist.utl.pt}
\affiliation{CFTP, Departamento de F\'{\i}sica, Instituto Superior T\'ecnico,
Av. Rovisco Pais, 1049-001 Lisboa, Portugal}
\author{
O. Oliveira}
%\email{orlando@teor.fis.uc.pt}
\affiliation{CFC, Departamento de F\'{\i}sica, Universidade de Coimbra, Rua Larga, 3004-516 Coimbra, Portugal}

\begin{abstract}
We study the long distance interaction for hybrid hadrons, with a static 
gluon, a quark and an antiquark with lattice QCD techniques. A Wilson loop 
adequate to the static hybrid three-body system is developed and, using a
$24^3 \times 48$ periodic lattice with $\beta=6.2$ and $a \sim 0.075$ fm, two 
different geometries for the gluon-quark segment and the gluon-antiquark 
segment are investigated.
When these segments are perpendicular, the static potential is 
compatible with confinement realized with a pair of fundamental strings, 
one linking the gluon to the quark and another linking the same gluon to the 
antiquark. When the segments are parallel and superposed, the total string 
tension is larger and agrees with the Casimir Scaling measured by Bali. This
can be interpreted with a type-II superconductor analogy for the 
confinement in QCD, with repulsion of the fundamental strings and with
the string tension of the first topological excitation of the string 
(the adjoint string) larger than the double of the fundamental string 
tension.
\end{abstract}

\maketitle

%================================================================
%================================================================
\section{Introduction}

Here we explore the static potential of the hybrid three-body system composed
of a gluon, a quark and an antiquark using lattice QCD methods. The Wilson 
loop method was deviced to extract from pure-gauge QCD the static potential 
for constituent quarks and to provide detailed information on the confinement 
in QCD. In what concerns gluon interactions, the first lattice studies were
performed by Michael 
\cite{Michael:1985ne,Campbell:1985kp}
and Bali extended them to other SU(3) representations 
\cite{Bali:2000un}.
Recently Okiharu and colleagues 
\cite{Okiharu:2004ve,Okiharu:2004wy}
extended the Wilson loop for tree-quark baryons
to tetraquarks and to pentaquarks. Our study of hybrids continues the lattice 
QCD mapping of the static potentials for exotic hadrons. 

The interest in hybrid three-body gluon-quark-antiquark systems is increasing 
in anticipation to the future experiments BESIII at IHEP in Beijin, GLUEX at 
JLab and PANDA at GSI in Darmstadt, dedicated to study the mass range of the 
charmonium, with a focus in its plausible hybrid excitations. Moreover, 
several evidences of a gluon effective mass of 600-1000 MeV from the Lattice 
QCD gluon propagator in Landau gauge, 
\cite{Leinweber,Oliveira_propagator}, 
from Schwinger-Dyson and Bogoliubov-Valatin solutions for the gluon propagator in
Landau gauge 
\cite{Fischer:2002eq},
from the analogy of confinement in QCD to 
supercondutivity 
\cite{Nielsen:1973cs},
from the lattice QCD breaking of the adjoint string 
\cite{Michael:1985ne},
from the lattice QCD gluonic 
excitations of the fundamental string 
\cite{Griffiths:1983ah}
from constituent gluon models 
\cite{Hou:1982dy,Szczepaniak:1995cw,Abreu:2005uw}
compatible with the lattice QCD glueball spectra 
\cite{glulat1,glulat2,glulat3,glulat4}, 
and with the Pomeron trajectory for high energy scattering 
\cite{Llanes-Estrada:2000jw,Meyer:2004jc}
may be suggesting that the static interaction for gluons is relevant.

\begin{figure}
\begin{picture}(350,90)(0,0)
\put(-20,-20){\includegraphics[width=0.7\textwidth]{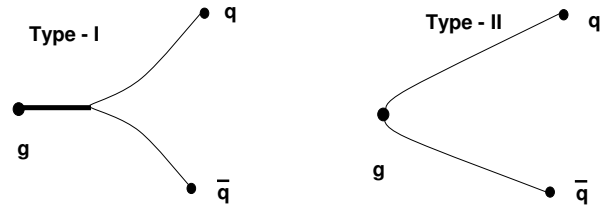}}
% Truque: Em word .doc ou power point .ppt fazer um grafico com autoshapes
% + ter cuidado para ficar em baixo a esquerda e nao muito grande
% + convert to pdf com o print para cutpdf ou pdf 
% OR: for small space + open with gsview +convert to eps 
% OR: for better resolution oped with adobe pdf + save as eps
%+ usar o includegraphics e nao o epsffile
\end{picture}
\caption{String attraction and fusion, and string repulsion, respectively
in type I and II superconductors} 
\label{typeI-II}
\end{figure}

Importantly, an open question has been residing in the potential for hybrid 
system, where the gluon is a colour octet, and where the quark and antiquark 
are combined to produce a second colour octet. While the constituent quark 
(antiquark) is usually assumed to couple to a fundamental string, in constituent
gluon models the constituent gluon is usually assumed to couple to an adjoint string. 
Notice that in lattice QCD, using the adjoint representation of SU(3), Bali 
\cite{Bali:2000un} 
found that the adjoint string is compatible with
the Casimir scaling, were the Casimir invariant $\lambda_i \cdot \lambda_j$ 
produces a factor of $9/4$ from the $q \bar q$ interaction to the $gg$ 
interaction. Thus we already know that the string tension, or energy per unit 
lenght, of the adjoint string is $1.125$ times larger than the sum of the 
string tension of two fundamental strings. How can these two pictures, of one 
adjoint string and of two fundamental strings, with different total string 
tensions, match? This question is also related to the superconductivity model 
for confinement, is QCD similar to a Type-I or Type-II superconductor?
Notice that in type Type-II superconductors the flux tubes repel each
other while in Type-I superconductors they attract each other and tend to fuse
in excited vortices. This is sketched in Fig. \ref{typeI-II}. The 
understanding of the hybrid potential will answer these questions.

In Section II we produce a Wilson Loop adequate to study the
static hybrid potential. In Section III we present the results of our
Monte-Carlo simulation, in a $24^3 \times 48$ 
pure gauge lattice for $\beta = 6.2$, 
corresponding to a lattice size of $(1.74 $ fm$ )^3 \times (3.48 $ fm$)$, assuming a 
string tension $\sqrt{\sigma} = 440$ MeV. In Section IV we interpret the 
results and conclude.

%================================================================
%================================================================
\section{Hybrid Wilson Loop}

\begin{figure}
\begin{picture}(350,140)(0,0)
\put(-20,-30){\includegraphics[width=1.0\textwidth]{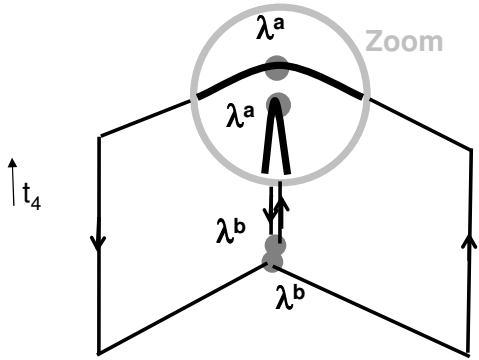}}
% Truque: Em word .doc ou power point .ppt fazer um grafico com autoshapes
% + ter cuidado para ficar em baixo a esquerda e nao muito grande
% + convert to pdf com o print para cutpdf ou pdf 
% OR: for small space + open with gsview +convert to eps 
% OR: for better resolution oped with adobe pdf + save as eps
%+ usar o includegraphics e nao o epsffile
\end{picture}
\caption{Wilson loop for the gqq potential} 
\label{Wilson Orlando}
\end{figure}

In principle any Wilson loop with a geometry similar to the one in Fig. 
\ref{Wilson Orlando} and describing correctly the quantum numbers of the hybrid 
is adequate, although the signal to noise ratio may depend in the choice of the
Wilson loop. A correct Wilson loop must include a SU(3) octet, the gluon, a 
SU(3) triplet, the quark and a SU(3) anti-triplet, the antiquark. It must also 
include the connection between the three links of the gluon, the quark and the
antiquark.

In the limit of infinite quark mass, a nonrelativistic potential $V$ can be
derived from the large time behaviour of euclidean time propagators. 
Typically, one has a meson operator $\mathcal{O}$ and computes the Green
function,
\begin{equation}
   \langle 0| \, \mathcal{O} (t) \, \mathcal{O} (0) \, | 0 \rangle ~
  \longrightarrow ~  \exp \{ - V t \}
\label{Green function}
\end{equation}
for large $t$. Different types of operators allow the definition of different
potentials. In the static gluon-quark-antiquark interaction, the static gluon 
can be replaced by a static quark-antiquark pair in a colour octet 
representation. In this way, we can construct the gluon-quark-antiquark Wilson
loop starting from the mesonic operator, 
\begin{equation}
\mathcal{O} (x) ~ = ~ \frac{1}{4} \, 
       \Big[ {\overline q} (x) \,  \lambda^a \, \Gamma_1 \, q(x) \Big]
       \Big[ {\overline q} (x) \,  \lambda^a \, \Gamma_2 \, q(x) \Big]
  \, ,
\label{mesonmeson}
\end{equation} 
where $\Gamma_i$ are spinor matrices. Using the lattice links to comply with 
gauge invariance, the second operator in eq. (\ref{mesonmeson}) can be made 
nonlocal to separate the quark and the antiquark from the gluon,
\begin{eqnarray}
 \mathcal{O} (x) & = & \frac{1}{4}
  \Big[ {\overline q} (x) \,  \lambda^a \, \Gamma_1 \, q(x) \Big]
\nonumber \\
   &  & 
 \Big[ {\overline q}(x - r_1 \hat{\mu}_1) 
             U_{\mu_1} (x - r_1 \hat{\mu}_1) \cdots U_{\mu_1} (x - \hat{\mu}_1)
\nonumber \\
   &  &   \lambda^a ~ \Gamma_2  \ \  U_{\mu_2} (x) \cdots U_{\mu_2} (x + (r_2-1)\hat{\mu}_2)
\nonumber \\
   &  &  
         q (x + r_2 \hat{\mu}_2 ) \Big] \, .
\label{op88}
\end{eqnarray}
The nonrelativisit potential requires the computation of the Green functions 
present in eq. (\ref{Green function}). Assuming that all quarks are of 
different nature, the contraction of the quark field operators gives rise to 
the gluon operator,
\begin{eqnarray}
W_{O}= &&
\frac{1}{16} 
\mbox{Tr} \Big\{ 
  U^\dagger_4 (t-1,x) \cdots U^\dagger_4 (0,x)  ~  \lambda^b 
\nonumber \\
  &&
   U_4 (0,x) \cdots U_4 (t-1,x)  ~ \lambda^a \Big\} ~ \times
\nonumber \\
\mbox{Tr} 
\Big\{ &  &
   U_{\mu_2} (t,x) \cdots U_{\mu_2} (t,x+(r_2-1)\hat{\mu}_2) 
\nonumber \\
  &&
   U^\dagger_4 (t-1,x + r_2 \hat{\mu}_2) \cdots 
                 U^\dagger_4 (0,x+r_2\hat{\mu}_2) 
\nonumber \\
  &&
   U^\dagger_{\mu_2} (0,x + (r_2-1) \hat{\mu}_2) \cdots 
                 U^\dagger_{\mu_2} (0,x) ~  \lambda^b 
\nonumber \\
  & &
   U^\dagger_{\mu_1} (0,x - \hat{\mu}_1) \cdots 
                 U^\dagger_{\mu_1} (0,x - r_1 \hat{\mu}_1 )
\nonumber \\
  &&
   U_4 (0,x - r_1 \hat{\mu}_1) \cdots 
                 U_4 (t-1,x-r_1\hat{\mu}_1)
\nonumber \\
  &&
   U_{\mu_1} (t,x - r_1 \hat{\mu}_1) \cdots 
                 U_{\mu_1} (t,x - \hat{\mu}_1 ) ~ \lambda^a \Big\} ~ .
\label{glue88}
\end{eqnarray}
Gauge invariance of (\ref{glue88}) can be proven with the help of the 
relation
\begin{equation}
\sum_a \, \left( \frac{\lambda^a}{2} \right)_{ij} \, 
          \left( \frac{\lambda^a}{2} \right)_{kl} ~ = ~
    \frac{1}{2} \delta_{il} \delta_{jk} - 
     \frac{1}{6} \delta_{ij} \delta_{kl} \, .
\label{exchange+identity}
\end{equation}

How does our operator relate with the operators used so far to investigate
the gluon interactions on the lattice? The gluonic time-like links used by Michael 
and collegues
\cite{Campbell:1985kp,Foster:1998wu}
to study the glue lump are the real $8 \times 8$
matrices,
\begin{equation} 
   U_4^{{\rm Adj \ }\alpha \beta} = 
   {1 \over 2} {\rm Tr}\{ U_4 {\lambda^\alpha}U_4^{\dagger}{\lambda^\beta} \} 
\label{Michael link}
\end{equation}
built from the usual SU(3) fundamental representation links $U_i$,
whereas in the investigation of Casimir scaling by Bali 
\cite{Bali:2000un}, 
the author worked directly with adjoint links, i.e. with the $8 \times 8$ matrix SU(3) 
representation. If one now 
compares the Wilson loop in eq. (\ref{glue88}) with eq. (\ref{Michael link}), 
it comes that when $t$ corresponds to a single lattice spacing, then the gluonic
trace, i. e. the first trace, of eq. (\ref{glue88}) is a ``Michael link''.

%================================================================
%================================================================
\section{The Static Hybrid Potential}

In this paper we consider two possible hybrid geometries: $\perp $ with the
quark-gluon segment perpendicular to the antiquark-gluon segment;
$\parallel$ with the quark-gluon segment parallel to the antiquark-gluon 
segment. We denote the potentials, respectively, $V_\perp ( r_1, r_2)$
and $V_\parallel ( r_1, r_2)$, where $r_1$ ($r_2$) is the quark-gluon
(antiquark-gluon) distance in lattice units, defined in eq. (\ref{glue88}).

Here we discuss the results of our simulation with 142
$24^3\times 48$, $\beta = 6.2$ pure-gauge Wilson action SU(3) configurations.
The configurations are generated with the version 6 of the MILC code \cite{Gauge},
via a combination of Cabbibo-Mariani and overrelaxed updates. In order to 
improve the signal to noise ratio, the links are replaced by ``fat links'' 
\cite{Blum} 
\begin{equation}
	U_{\mu}( \mathbf{s} ) \rightarrow
	\frac{1}{ 1 + 6 w }
	\big(
	U_{\mu}( \mathbf{s} )
	+ w
	\sum_{\mu \neq \nu} U_{\nu}( \mathbf{s} ) U_{\mu}(\mathbf{s} + \nu)  U_{\nu}^\dagger( \mathbf{s} + \mu  )
	$$$$
	+ U_{\nu}^\dagger( \mathbf{s} - \nu ) U_{\mu}(\mathbf{s} - \nu)  U_{\nu}( \mathbf{s} + \mu - \nu  )
	\big)
\end{equation}
followed by a projection into SU(3). We use $w = 0.2$ and iterate this 
''fat link'' $25$ times both in the spatial direction 
and in the temporal one. The temporal smearing reduces the short-range
Coulomb potential but produces a clearer signal for the long-range potential,
the one we are interested in.
Furthermore, to improve the quality of the
signal, we explore the symmetry $r_1 \leftrightarrow r_2$ when computing
$V_\perp(r_1,r_2)$ and $V_\parallel(r_1,r_2)$.

\begin{figure}[t]
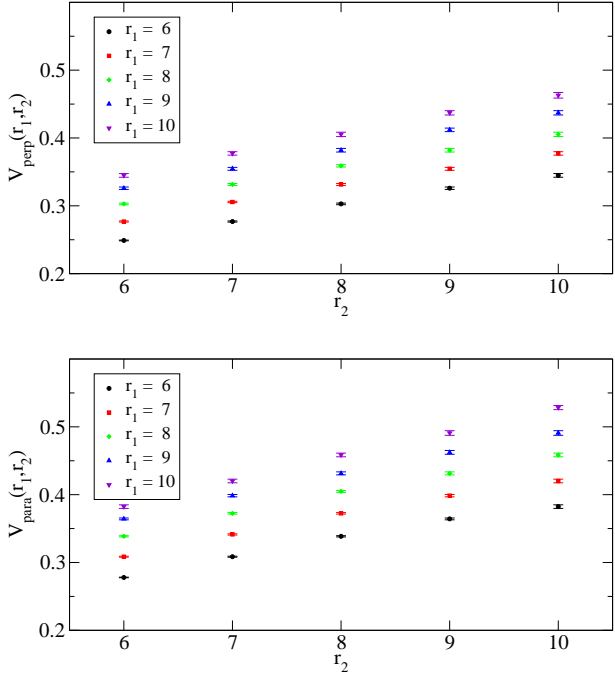

\begin{picture}(230,250)(0,0)
\put(0,-5){ \includegraphics[width=0.45\textwidth]{VP_a.eps} }
\put(0,130){ \includegraphics[width=0.45\textwidth]{VL_a.eps} }
\end{picture}
\caption{Potential in lattice spacing units, for the system $q{\overline q}g$, 
respectively for the $\perp$ and $\parallel$ geometry 
}
\label{potPerpandPara}
\end{figure}

Using eq. (\ref{Green function}), the static potentials are extracted
from the fit of minus the $\log $ of the Wilson loop, $-\log W$,
for large euclidian time $t$.
This fit provides us with the potential, and we estimate the respective error bar with the jackknife method.

We are essentially interested in the largest possible distances,
to compare the different possible string tensions. With
$24^3 \times 48$ periodic lattices, the maximum distance we reach
is $10 \, a$. In this way we avoid possible finite volume
effects. At $12 \,a $, due to our periodic boundary conditions, 
we already could approach a maximum, deviating from the linear behaviour.
Notice that we calibrate our lattice spacing $a \sim 0.075 $ fm as in 
Bali and Schilling
\cite{Bali:1993ab}. Thus our maximal distance
is still comfortably shorter than the string breaking distance, larger than
1 fm, and comfortably longer than the pertubative distance of say,  0.3 fm.
We also start measuring the potentials at the distance of $6 \, a$ because
we are interested in studying the long-distance, non-pertubative part of
the static potentials. Our results for the static hybrid potentials 
$V_\perp$ and $V_\parallel$ are displayed in Fig. \ref{potPerpandPara}.

Again, to get the string tensions $\sigma$, we fit the large distance part
of the potentials with a linear potential, with the same method
we used for the temporal fit to extract the static potentials.
We admitting a singlet quark-antiquark singlet string tension of
$440$ MeV, which corresponds to an inverse lattice spacing of 
${a}^{-1} = 2718 \pm 32$ MeV, according to Bali and Schilling 
\cite{Bali:1993ab}.

\begin{figure}[t]
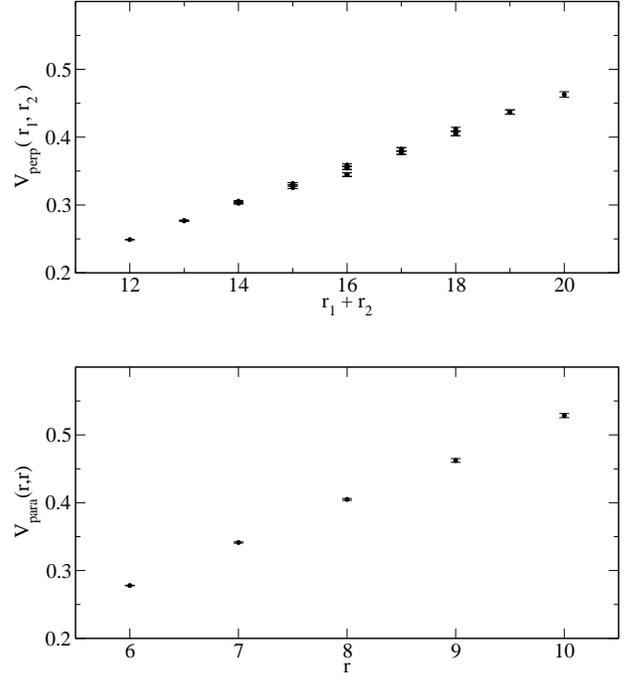

\begin{picture}(230,250)(0,0)
\put(0,-5){ \includegraphics[width=0.45\textwidth]{VRR_a.eps} }
\put(0,130){ \includegraphics[width=0.45\textwidth]{V+_a.eps} }
\end{picture}
\caption{
Potential for the system $q{\overline q}g$, in lattice spacing units,
respectively for the $\perp$  geometry as a function of $r_1 + r_2$
and for the $\parallel$ geometry as a function of $r_1 = r_2 = r$ 
} 
\label{potparaRRSumperp}
\end{figure}

To study the onset of two fundamental strings, we plot in Fig. \ref{potparaRRSumperp}
the perpendicular geometry potential $V_\perp$ as a function of the sum of the two
distances in lattice spacing units, $r_1$ between the quark and the gluon and 
$r_2$ between the antiquark and the gluon, as in eq. (\ref{glue88}). Indeed
the potential is linear in the sum of the distances. 
We further fit the potential to $V_\perp(r_1,r_2) = c_0 + \sigma (r_1 + r_2)$, and get
\begin{equation}
	\sqrt{\sigma} = 441 \pm 6 \, \text{MeV} = ( 1.00 \pm 0.01 ) \sqrt{\sigma_0}
\end{equation}
$( \chi^2/dof = 1.34802 )$
which is consistent with $ \sigma = \sigma_0 $, reinforcing the picture that, at long 
distances we have two fundamental strings one linking the quark and the gluon and, the other,
linking the antiquark to the gluon.

To study if the double fundamental string picture can also be compatible with the Casimir
scaling result found by Bali
\cite{Bali:2000un},
we also consider the case where the quark and antiquark are superposed. In this case
the static quark and antiquark are equivalent to a static gluon, and therefore
our potential is equivalent to a static gluon-gluon potential. This is the case
of the parallel geometry potential $V_\parallel$ when the two distances 
$r_1$ between the quark and the gluon and 
$r_2$ between the antiquark and the gluon, as in eq. (\ref{glue88}), are identical,
$r_1=r_2$. This is plotted in Fig. \ref{potparaRRSumperp} and indeed we find a linear
behaviour.
We further fit the static potential of the parallel geometry with $r_1 = r_2 = r$,
by $V_\parallel(r,r) = c_0 + \sigma r$, and we get
\begin{equation}
	\sqrt{\sigma} = 681 \pm 9 \, \text{MeV} = (1.55 \pm 0.02) \sqrt{\sigma_0}
\end{equation}
$( \chi^2 / dof = 0.235172 )$ wich is a little larger than the Casimir scaling 
$\sqrt{\sigma} = \frac{3}{2} \sqrt{\sigma_0}$ ratio. Nevertheless our results
confirm with the ones of Bali
\cite{Bali:2000un}
since our error bars match. 
The increase or the static hybrid potential when the quark and antiquark 
are superposed can be interpreted with a repulsive energy between the two
fundamental strings. This repulsive energy also exists in Type-II superconductors.

%================================================================
%================================================================
\section{Conclusion}

We explore, in
$24^3 \times 48$ periodic lattices with $\beta=6.2$ and $a \sim 0.075$, two 
different geometries for the gluon-quark segment and the gluon-antiquark 
segment. When these segments are perpendicular, the static potential is 
consistent with confinement realized with a pair of fundamental strings, 
one linking the gluon to the quark linking the same gluon to the 
antiquark. When the segments are parallel and superposed, the total string 
tension is larger and is compatible with a repulsive energy between the two
fundamental strings. Notice that when the two segments are parallel and
superposed, the total string tension is also compatible with the Casimir 
Scaling measured by Bali. 

This can be interpreted with a type-II superconductor analogy for the 
confinement in QCD, with repulsion of the fundamental strings and with
the string tension of the first topological excitation of the string 
(the adjoint string) larger than the double of the fundamental string 
tension. Nevertheless, because the energy of two fundamental strings plus 
the repulsive energy measured here is quite similar to the energy of the 
adjoint string measured by Bali \cite{Bali:2000un},
this shows that the pure gauge QCD is similar to a Type-II superconductor
quite close to the phase transition to a Type-I superconductor.

Our results are important for constituent models for hybrids and glueballs.
In the three-body hybrid, with one quark, one antiquark and one gluon,
our results suggest that the best potential model has only two
fundamental strings, plus a repulsion acting only when the two fundamental strings 
are close. In the two body gluon-gluon glueball, our results suggest that
the string tension is similar to the one of the Casimir Scaling model, 
with a factor of the order of $9 \over 4$ when compared with the quark-antiquark
potential. We can also extrapolate our result for three-body glueballs, relevant
for the odderon problem. With three gluons, a triangle formed by three fundamental
strings costs less energy than three adjoint strings with a starfish-like geometry.
Thus we anticipate that the three-gluon potential is similar to a sum 
of three mesonic quark-antiquark interactions, plus a repulsion acting only when 
there is superposition of the fundamental strings.

%================================================================
%================================================================
\acknowledgements

This work was financed by the FCT contracts POCI/FP/63436/2005
and POCI/FP/63923/2005.

%================================================================
%================================================================

%
\end{document}